\documentclass[useAMS]{mn2e}
\def\gtrsim{\mathrel{\hbox{\rlap{\hbox{\lower4pt\hbox{$\sim$}}}\hbox{$>$}}}}
\usepackage{graphicx}  
\usepackage{longtable}                
\usepackage{subfig}
\usepackage{caption}
\usepackage{everysel}
\usepackage{keyval}
\usepackage{ragged2e}
\usepackage{rotating}
\usepackage{hyperref}

\usepackage{fixltx2e}

\title[Fomalhaut b as neutron star]{The companion candidate near Fomalhaut - a background neutron star?}

\author[R. Neuh\"auser et al.]{R. Neuh\"auser$^{1}$\thanks{rne@astro.uni-jena.de},
M.M. Hohle$^{1,2}$\thanks{hohle@lmb.uni-muenchen.de}, 
C. Ginski$^{1,3}$\thanks{ginski@strw.leidenuniv.nl}, 
J.G. Schmidt$^{1}$\thanks{schmidt.janos@uni-jena.de},
\newauthor V.V. Hambaryan$^{1}$\thanks{vvh@astro.uni-jena.de}, 
and T.O.B. Schmidt$^{1,4}$\thanks{tschmidt@hs.uni-hamburg.de}\\
$^{1}$ Astrophysikalisches Institut und Universit\"ats-Sternwarte Jena, Schillerg\"asschen 2-3, 07745 Jena, Germany \\
$^{2}$ Graduate School of Quantitative Biosciences Munich, Genecenter of the LMU, Feodor-Lynen-Str. 25, 81377 Munich, Germany \\
$^{3}$ Leiden Observatory, Leiden University, P.O. Box 9513, 2300 RA Leiden, The Netherlands \\
$^{4}$ Hamburger Sternwarte, Gojenbergsweg 112, 21029 Hamburg, Germany}

\begin{document}

\bibliographystyle{mn2e}

\date{Accepted 2014 Dec 27. Received 2014 Dec 18; in original form 2014 Jun 26}

\pagerange{\pageref{firstpage}--\pageref{lastpage}} \pubyear{2002}

\maketitle

\label{firstpage}

\begin{abstract}
The directly detected planetary mass companion candidate close to the young, nearby star Fomalhaut is a subject of intense discussion. 
While the detection of common proper motion led to the interpretation as Jovian-mass companion, later non-detections in the infrared raised doubts. 
Recent astrometric measurements indicate a belt crossing or highly eccentric orbit for the object, if a companion, 
making the planetary interpretation potentially even more problematic.\\ 
In this study we discuss the possibility of Fomalhaut\,b being a background object with a high proper motion. 
By analysing the available photometric and astrometric data of the object, we show that they are fully
consistent with a neutron star: Neutron stars are faint, hot (blue), and fast moving.
Neutron stars with 
an effective temperature of the whole surface area being
112,000 K to 126,500 K (with small to negligible extinction)
at a distance of roughly 11 pc (best fit) would be consistent with all observables, namely with the
photometric detections in the optical, with the upper limits in the infrared and X-rays,
as well as with the astrometry (consistent with a distances of 11 pc or more and high proper motion as typical for neutron stars) 
as well as with non-detection of pulsation (not beamed).
We consider the probability of finding an unrelated object or even a neutron star nearby and mostly co-aligned in proper motion with Fomalhaut\,A 
and come to the conclusion that this is definitely well possible.
\end{abstract}


\section{Introduction}

The direct detection of a possibly planetary mass object near the 
star Fomalhaut by Kalas et al. (2008)
was widely regarded as a great success for the direct imaging detection method. 
The separation between Fomalhaut A and b is some 100 au or 13 arc sec.
In addition to this published planetary mass companion candidate (called Fomalhaut b), 
Fomalhaut A (the central star) is surrounded by a well resolved dust belt, 
which was most recently studied with Herschel (Acke et al. 2012) and ALMA (Boley et al. 2012). 
The projected position of the tentative companion was interpreted to indicate that it had cleared the gap in this belt. 
The presence of the belt close to the companion candidate constrained the upper mass 
limit of the companion candidate to a few Jupiter masses (Kalas et al. 2008).

The star Fomalhaut\footnote{This star is also called $\alpha$ PsA,
i.e. the brightest star in the Southern Fish, the name
{\it Fomalhaut} comes from the Arabic {\it fam al-\d{h}\={u}t al-jan\={u}b\={\i}}
meaning {\it mouth of the southern fish}, (Kunitzsch \& Smart 1986).}
(Fomalhaut A) has the following relevant properties (all for the star A):
\begin{itemize}
\item Position J2000.0: $\alpha$ = 22h 57m 39s and $\delta = -29^{\circ}~37^{\prime}~20^{\prime \prime}$ (Hipparcos, van Leeuwen 2007).
\item The distance as measured by Hipparcos is $7.70 \pm 0.03$ pc (van Leeuwen 2007).
\item Proper motion as also measured by Hipparcos is $\mu _{\alpha} = 328.95 \pm 0.50$ mas/yr
and $\mu _{\delta} = -164.67 \pm 0.35$ mas/yr (van Leeuwen 2007).
\item The spectral type is A4V as obtained by an optical spectrum (Gray et al. 2006);
given this spectral type, the color index is close to zero, e.g. B-V=0.09 mag (e.g. Ducati 2002).
\item The optical brightness is V = 1.16 mag (e.g. Ducati 2002).
\item The age was recently determined to be $440 \pm 40$ Myr by kinematic membership to
the young Castor Moving Group (Barrado y Navascues 1998, Mamajek 2012).
\end{itemize}

The companion was originally discovered in the optical bands of the Hubble Space Telescopes (HST) Advanced Camera for Survey (ACS, Ford et al. 1998). 
However, several attempts to detect the object in the near and mid infrared (see e.g. Kalas et al. 2008 and Janson et al. 2012) failed (see Table 1). 
This was most troublesome, given that a (few hundred Myr) young cooling Jovian-mass object should be much brighter in the infrared than in the optical. 
Furthermore, the latest astrometric measurements by Kalas et al. (2013) indicate that the object would either cross the dust belt 
or that it would be on a highly eccentric orbit which is not in alignment with the belt at all. 
These two facts together have prompted us to seek for an alternative explanation which might explain all the observations and 
finally resolve some of the apparent contradictions. In the following we will first briefly discuss the various scenarios that have been proposed so far and will then present our own considerations.

In this paper, we first review the observations of Fomalhaut b (Sect. 2) and the interpretations
as Jupiter-mass planet (Sect. 2.1), as super-Earth (Sect. 2.2), and as dust cloud (Sect. 2.3).
Then, we consider the background hypothesis as either a White Dwarf (Sect. 3.1) or neutron star (Sect. 3.2.);
we discuss the astrometry, the X-ray data, and the optical and IR photometry, to constrain the neutron star
properties (to be consistent with all observables). In Sect. 3.2.5, we also discuss the probability to
find a background object or even a neutron star close to a star like Fomalhaut. We conclude in Sect. 4.

\section{Fomalhaut~\lowercase{b} as a gravitationally bound object}

\subsection{Fomalhaut~b as a Jupiter-mass planet}

The first interpretation of the available data by Kalas et al. (2008) led to the conclusion that the object may be a giant planet. 
From stability considerations of the dust belt, 
Kalas et al. (2008) and Chiang et al. (2009) inferred that the mass of the object should be $\leq$ 3\,M$_{\rm J}$. 
Larger masses would lead to either smaller orbits than could be inferred from the astrometry, 
or higher belt eccentricities than are observed. 
Kalas et al. (2008) concluded that if the flux in the optical wavelength range originates in the photosphere of a cooling planet, 
then the object needs to be cooler than 400\,K.  Otherwise too much flux would be produced at 1.6\,$\mu$m. 
They suspected that their non-detections at 1.6\,$\mu$m and 3.8\,$\mu$m might be due to model uncertainties. 
However, Marengo et al. (2009) and Janson et al. (2012) present Spitzer IRAC upper detection limits at 4.5\,$\mu$m, 
which puts additional constraints on the mass of a possible giant planet.  
Janson et al. (2012) conclude that the optical flux cannot stem from a planet's photosphere, 
especially since the flux at 0.6\,$\mu$m is 20 to 40 times brighter than expected for an object with $\sim$3\,M$_J$ and a few 
hundred Myr (Fortney et al. 2008, Burrows et al. 2003). 
On the contrary, Currie et al. (2012) and Galicher et al. (2013) argue that an 0.5-1\,M$_J$ object would not have been detected at 4.5\,$\mu$m 
(using models by Spiegel \& Burrows 2012 and Baraffe et al. 2003). 
In addition, Galicher et al. (2013) present upper detection limits at 1.1\,$\mu$m, 
which are consistent with this upper mass limit. 
However, there is a general agreement in all aforementioned studies that the flux in the optical wavelength range cannot stem completely (or at all) from a planet's photosphere.

In addition to the discussed over-luminosity in the optical wavelength range, Kalas et al. (2008) and Janson et al. (2012) report significant (5-8 $\sigma$) variability 
of the flux at 0.6\,$\mu$m, which could not be explained by a thermal emission from a planet's photosphere. 
Kalas et al. (2008) propose that there might be a 20-40\,R$_J$ accretion disk around the assumed planet. 
The disk would reflect light from the primary star, which explains the optical excess flux. 
Furthermore, they argue that the variability could then be explained by accretion driven H$_\alpha$ emission. 
Janson et al. (2012) strongly disagree, stating that at the high system age moons should have formed in a possible accretion disk, 
and thus the reflective surface should be reduced. 
Also, they think it is unlikely that accretion driven H$\alpha$ emission can explain the variability, 
because the accretion rate would have to be similar to young T Tauri stars.  
Currie et al. (2012) and Galicher et al. (2013) re-analyzed the same optical data and did not detect any significant variability at 0.6\,$\mu$m. 
Thus, they are not excluding a dust disk around a Jupiter-mass planet.

The most recent study by Kalas et al. (2013) incorporates new astrometric measurements taken with the HST STIS (Space Telescope Imaging Spectrograph, Woodgate et al. 1998) in 2010 and 2012. 
They find that the object is most likely on a ring-crossing orbit with a high semimajor axis and eccentricity. 
One explanation for that, assuming Fomalhaut b is a planet, would be that it had a close encounter with a further-in massive planet 
and was scattered out. 
However, Kenworthy et al. (2013) performed deep coronagraphic imaging and can rule out further-in objects with 12-20\,M$_J$ at 4-10\,au. 
They state that this effectively rules out scattering scenarios, which makes the orbit elements recovered by Kalas et al. (2013) somewhat peculiar.

Given that five astrometric data points are available only for four different epochs separated by a few years, 
any orbit fits with periods of hundreds of years (Kalas et al. 2013) suffer from high uncertainties anyway.  
Given the large separation between the two objects (star and presumable planet), even if bound, 
other planet detection techniques like the radial velocity or transit technique cannot be applied.

Given the various and sometimes contradictory arguments, 
the existence of a giant planet at the position of Fomalhaut b is still possible, but seems increasingly problematic.

\begin{table*}
  \caption{Photometric observation epochs and analysis by various authors of Fomalhaut b.}
	\label{tab: obs}
  \begin{tabular}{lllllccc}
  \hline
      
Date	& Telescope & Instrument & Filter &  Ref. & Det. ? & app. magnitude /  & flux \\ 
        &           &	         &	  &       &        & upper limit [mag] & [erg/cm$^{2}$/s/\AA ]	\\ \hline
 2004 Sep. 26						& HST							&	ACS/HRC					& F606W													&	Currie et al. 2012			& yes										& $24.92 \pm 0.10$ & $3.14 \pm 0.29 \cdot 10^{-19}$ \\
 2004 Oct. 25						& HST							&	ACS/HRC					& F606W													&	Kalas et al. 2008			& yes										& $24.43 \pm 0.09$ & $4.93 \pm 0.41 \cdot 10^{-19}$ \\
 2004 Oct. 26						& HST							&	ACS/HRC					& F606W													&	Kalas et al. 2008			& yes										& $24.29 \pm 0.08$ & $5.61 \pm 0.41 \cdot 10^{-19}$ \\

 2005 July 21						& KeckII					&	NIRC2						& H															&	Kalas et al. 2008		& no										& $\ge 22.9$ & $\le 8.28 \cdot 10^{-20}$ \\
 2005 Oct. 21						& KeckII					&	NIRC2						& CH$_4$S												&	Kalas et al. 2008			& no										& $\ge 20.6$ & \\

 2006 July 14-20				& HST							&	ACS/HRC					& F435W													&	Kalas et al. 2008			& no										& $\ge 24.7$ &  $\le 8.36 \cdot 10^{-19}$ \\
 2006 July 14-20				& HST							&	ACS/HRC					& F435W													&	 Currie et al. 2012 			& yes										& $25.22 \pm 0.18$ & $5.18 \pm 0.86 \cdot 10^{-19}$ \\
 2006 July 14-20				& HST							&	ACS/HRC					& F606W													&	Kalas et al. 2008			& yes										& $25.13 \pm 0.09$ & $2.59 \pm 0.21 \cdot 10^{-19}$ \\
 2006 July 14-20				& HST							&	ACS/HRC					& F606W													&	 Currie et al. 2012 			& yes										& $24.97 \pm 0.09$ & $3.00 \pm 0.25 \cdot 10^{-19}$ \\
 
 2006 July 14-20				& HST							&	ACS/HRC					& F814W													&	Kalas et al. 2008			& yes										& $24.55 \pm 0.13$ & $1.69 \pm 0.20 \cdot 10^{-19}$ \\
 2006 July 14-20				& HST							&	ACS/HRC					& F814W													&	 Currie et al. 2012 			& yes										& $24.91 \pm 0.20$ & $1.21 \pm 0.22 \cdot 10^{-19}$ \\

 2008 Sep. 17-18				& Gemini North		&	NIRI						& L'			& Kalas et al. 2008		& no										& $\ge 16.6$	& $\le 1.22 \cdot 10^{-18}$ \\
 2009 Aug. 16						& Subaru					&	IRCS						& J															&	 Currie et al. 2012 			& no										& $\ge 22.22$ & $\le 4.01 \cdot 10^{-19}$ \\

 2010 Aug. 8 -					& Spitzer					&	IRAC						& 4.5$\mu$m											&	Janson et al. 2012		& no										& $\ge 16.7$		& $\le 5.71 \cdot 10^{-19}$ \\
 2011 Jul. 23 					& 								&									&																&																	& 											& 				& \\

\hline
\end{tabular}

\end{table*}

\subsection{Fomalhaut~b as a super-Earth}

If there is a central object associated with the source Fomalhaut b, 
then Janson et al. (2012) state that its mass should be limited to $\leq$10\,M$_{Earth}$ if there is a ring-crossing orbit, which the astrometry suggests. 
This is in order to prevent the object from significantly influencing the observed belt geometry (see also Kennedy \& Wyatt 2011). 
For such an object to exhibit the observed fluxes in the optical wavelenth range, 
the object would need to be significantly hotter than a cooling planet of that mass. 
Janson et al. (2012) propose a scenario where the object had undergone an intense bombardment of planetesimals within the time scale of $\sim$10$^4$yr. 
However, they recognize that the observation of such an event seems improbable due to the short time scale, 
as compared to the age of the system. In addition, this scenario is not entirely compatible with the high flux at 0.6$\mu$m.

Another scenario proposed by Janson et al. (2012) is a 10\,M$_{Earth}$ object with a cloud of planetessimals which 
are producing the dust that reflects the starlight. 
However, this scenario would not explain the aforementioned variability at 0.6$\mu$m which was detected by two independent studies. 
It is also questionable why the orbit of such an object would exhibit a high semi-major axis and 
eccentricity as found by Kalas et al. (2013) if scattering scenarios can be ruled out (Kenworthy et al. 2013).

In general, while a smaller planetary mass object does not exhibit the same problems with the infrared detection limits as a more massive object, 
the orbit of such a low mass companion still seems peculiar. In addition, it is still challenging to explain the optical flux in such a scenario. 

\subsection{Fomalhaut~b as a dust cloud}
\label{dust-cloud}

Kalas et al. (2008) originally discussed the possibility that there might not be a central object associated with the source Fomalhaut~b, 
but that it is rather a dust cloud produced by the recent collision of two planetesimals. 
They reject this possibility because they think it is improbable to observe such a collision at the location of Fomalhaut~b, 
due to the low density of planetessimals outside the dust belt. In addition, 
they state that a dust cloud would not account for the variability at 0.6$\mu$m. 
Janson et al. (2012) note that the observation of such a dust cloud might not be as improbable as Kalas et al. (2008) state. 
They argue that such collisions should indeed happen more frequently inside the dust belt, 
but are not observable at this location due to the speckle-like nature of such sources.  
Thus the probability of observing one such collision outside the dust belt is not negligible. 
Galicher et al. (2013) argue that their detection of Fomalhaut b at 0.4$\mu$m would fit well with a dust cloud younger than 500\,yr composed of water ice 
or refractory carbonatious small grains, as originally proposed by Kalas et al. (2008). 
Furthermore, Galicher et al. (2013) do not detect the variability at 0.6$\mu$m, 
which was one of the main arguments against this scenario by Kalas et al. (2008). 
However, Currie et al, (2012) contend that the observation of an unbound dust cloud should be unlikely because 
Keplerian shear would spread out such a cloud. This small time frame as compared to the system age would make the observation of such a cloud implausible. 
We want to note that the study by Galicher et al. (2013) finds that the object Fomalhaut~b can be fitted slightly better with an extended source (0.58\,au) 
than with a point source.  This is, however, on a very low significance level, and other studies have not mentioned the possible resolved nature of the source.

Overall the dust cloud scenario may appear to be a possible scenario if Fomalhaut~b is extended and gravitationally bound to Fomalhaut A.

\section{Fomalhaut~\lowercase{b} as a background object}

When evaluating the background hypothesis compared to the possibility of a gravitational bound companion, usually a non-moving background object is assumed: 
One first tests the hypothesis of a bound companion with the null hypothesis that both the central star and the companion candidate have identical proper motion;
then one tests the background hypothesis with the null hypothesis that the central star has its finite (known) proper motion 
and that the companion candidate proper motion is zero.\footnote{Even if both objects have common proper motion, this is not yet a proof that
they orbit each other. Even the detection of curvature in the motion of the companion may not yet be a proof that they orbit each other,
if the curvature would also be consistent with a hyperbolic orbit of a recently ejected object. The detection of curvature in a form
that is not consistent with a hyperbolic orbit would be a proof that both objects orbit each other.}
However, this method may fail if the object of interest is a moving background object with a considerable proper motion. 
Therefore, we discuss the possibility that Fomalhaut~b could be a moving background object, unrelated to the primary star.

The HST magnitudes of Fomalhaut~b point to a flat SED (Kalas et al. 2008, Currie et al, 2012) that rules out any ordinary star given the faint magnitudes. 
Since the position of Fomalhaut~b changes only slightly over years with respect to the primary star, 
its proper motion should be roughly that of the primary star. Only a white dwarf or a neutron star are fast moving and dim enough to match these constrains.   

\subsection{Fomalhaut~b as a white dwarf}

The absolute visible magnitude of white dwarfs ranges from 10 to 15~mag (Wood \& Oswald 1998), Fomalhaut b has an apparent visible magnitude of $\sim 25$ mag (Table 1).
Even taking visual absorption caused by the ISM into account, 
the putative white dwarf would have a distance of at least 0.5~kpc, but 
up to 5~kpc (brightest white dwarf; no absorption). From these estimates, 
the object would have a projected spatial velocity (2D) of 900-9000~km/s (applying Fomalhaut's proper motion). 
In addition the putative white dwarf would have an unknown radial velocity component that would raise the spatial velocity (3D) to even larger values. 
These numbers are rather unrealistic, since the fastest known white dwarf moves 
with 450~km/s (Oppenheimer et al. 2001, Wood \& Oswald 1998). Hence, a white dwarf as a putative background object is very unlikely.

\begin{table}
\centering
\caption{Photometry of two brightest known isolated neutron stars as measured with HST (Kaplan et al. 2011)}
\begin{tabular}{ccc} \hline
wavelength    & width               & flux F$_{\lambda}$ \\ 
$[\AA ]$       &   [\AA ]            & [erg/cm$^{2}$/s/\AA ] \\ \hline
\multicolumn{3}{c}{RXJ~1856.4-3754} \\ \hline
1707.4        &     515.2           &    $1.50 \pm 0.13 \cdot 10^{-17}$ \\
2960.5        &     877.3           &    $2.34 \pm 0.14 \cdot 10^{-18}$ \\
4444.1        &     1210.5          &    $4.63 \pm 0.24 \cdot 10^{-19}$ \\
4739.1        &     1186.4          &    $3.84 \pm 0.50 \cdot 10^{-19}$ \\
5734.4        &     2178.2          &    $1.51 \pm 0.47 \cdot 10^{-19}$ \\ \hline
\multicolumn{3}{c}{RXJ~0720.4-3125} \\ \hline
1370          &     320             &    $7.94 \pm 1.34 \cdot 10^{-18}$ \\
1480         &      280             &    $5.92 \pm 0.57 \cdot 10^{-18}$ \\
2320         &      1010            &    $1.07 \pm 0.13 \cdot 10^{-18}$ \\
4739.1       &      1186.4          &    $1.13 \pm 1.45 \cdot 10^{-19}$ \\
5850         &      4410            &    $7.73 \pm 0.68 \cdot 10^{-20}$ \\ \hline
\end{tabular}
\end{table}

\subsection{Fomalhaut~b as a neutron star}

Young (1~Myr), hot (1~MK), and close-by ($\le 500$ pc) neutron stars (NSs) have visual magnitudes 
ranging from 25-27~mag (see e.g. Kaplan et al. 2011 for a compilation), i.e. as faint as Fomalhaut b, but more distant.
These objects are not necessarily radio pulsars and therefore Fomalhaut~b (if a NS) could have remained undetected. 
Indeed, the radio-quiet X-ray emitting NS RXJ~1856.4-3754,
the first such object not powered by rotation, has V=25.6 mag, and it was discovered 
by coincidence by Walter et al. (1996) as they actually searched for T Tauri stars.
Due to the supernova kick, the spatial velocities of NSs are on average much larger than those of 
e.g. white dwarfs and peak at $\sim 400$~km/s, the fastest known NS moves with 1500~km/s (Hobbs et al. 2005).

\subsubsection{Astrometry: Proper motion and parallax}

Since the companion candidate to Fomalhaut (i.e. object b) was found to be (at least nearly) co-moving to Fomalhaut A,
both Fomalhaut A and Fomalhaut b have very similar proper motions. We can therefore estimate the motion of Fomalhaut b by using the proper
motion of Fomalhaut A: $\mu_{\alpha} = 328.95 \pm 0.50$ mas/yr and $\mu_{\delta} = -164.67 \pm 0.35$ mas/yr (Hipparcos).
The proper motion of both Fomalhaut~A and b are then $\sim 368$~mas/yr.
This proper motion of Fomalhaut b would then be equivalent to a tangential (2D) velocity of $\sim 30$~km/s for 11~pc distance
(i.e. background to Fomalhaut A), or 170~km/s for 100~pc distance. Such velocities are fully consistent with NS velocities.

Since Fomalhaut A is a very nearby star (7.7 pc), its parallactic motion (wobble) is large (a parallax of 130 mas).
Since very precise astrometry for both Fomalhaut A and b are available (which were used to show their common proper motion),
one can check whether some differential parallactic motion between the star and the companion candidate are detectable:
If the presumable companion candidate would be in the distant background, one would not detect any significant parallactic
motion for the companion, but of course still large parallactic motion for the star Fomalhaut A.

To investigate whether the relative astrometry between Fomalhaut A and b would be consistent with Fomalhaut b 
being a background object, we tried to fit the data points with a differential proper motion and differential parallax. 
Results are shown in Fig. 1. In principle, our best fit in terms of reduced $\chi^2$ (0.16) yields a 
differential parallax of $39.1$\,mas and a differential proper motion of $-50.4$\,mas/yr in RA and $107.8$\,mas/yr in Dec. 
This corresponds to Fomalhaut b being in the background behind Fomalhaut A at a total distance of 11\,pc.
However, as indicated by the small reduced $\chi^2$, this result is not significant due to the low number of data 
points and their uncertainties. We also fit differential motion to the data points without any differential parallax. 
The resulting linear fit has only a marginally worse reduced $\chi^2$ (0.33) as compared to the best 
fit and is also fully consistent with all measurements. Finally, we repeated the same fitting procedure 
but with the maximum possible differential parallax of $129.8$\,mas. The resulting fit is consistent with all 
measurements but the one taken in 2010. 
However, this measurement has the highest uncertainty of all astrometric data points and 
lies still within two $\sigma$ of the fit and thus also for this scenario a reasonable reduced $\chi ^2$ of 1.41 was calculated.

This analysis shows that the astrometry is in principle compatible with Fomalhaut b being a background object at any distance (behind Fom A), 
although a distance of 11\,pc is slightly favored. However, as pointed out, we can not exclude that Fomalhaut b is at the same distance as Fomalhaut A.

Fomalhaut is located at a galactic latitude of $-64.9^{\circ}$, i.e. south of the Galactic plane.
The proper motion of both the star Fomalhaut A and its companion candidate (or the nearby neutron star)
are moving (in both equatorial and galactic coordinates) towards the south-south-east, i.e. away from the Galactic plane.
This would be consistent with a young neutron star which was recently born in the Galactic plane.
Since the Sun is currently $26 \pm 3$ pc north of the Galactic plane (Majaess et al. 2009),
and since Fomalhaut is $7.70 \pm 0.03$ pc away from the Sun (mostly towards the galactic south),
Fomalhaut (and its companion candidate) are currently $33.7 \pm 3$ pc south of the Galactic plane.
For the largest one-dimensional velocity known for a neutron star (1285 km/s for PSR B2011+38, Hobbs et al. 2005),
our object would have needed (at least) $2.6 \pm 0.2$ kyr to travel from the Galactic plane to its current position;
for the mean one-dimensional neutron star velocity ($133 \pm 8$ km/s, Hobbs et al. 2005),
it would have needed $248 \pm 37$ kyr.
Of course, it could have formed outside of the Galactic plane, or it may have oscillated around the plane
one or several times (and/or have orbited the Galactic center one or more times).
Its current position south of the Galactic plane together with its motion away from the plane
would be consistent with a young neutron star.

\begin{figure*}
\centering
\includegraphics[width=0.96\textwidth]{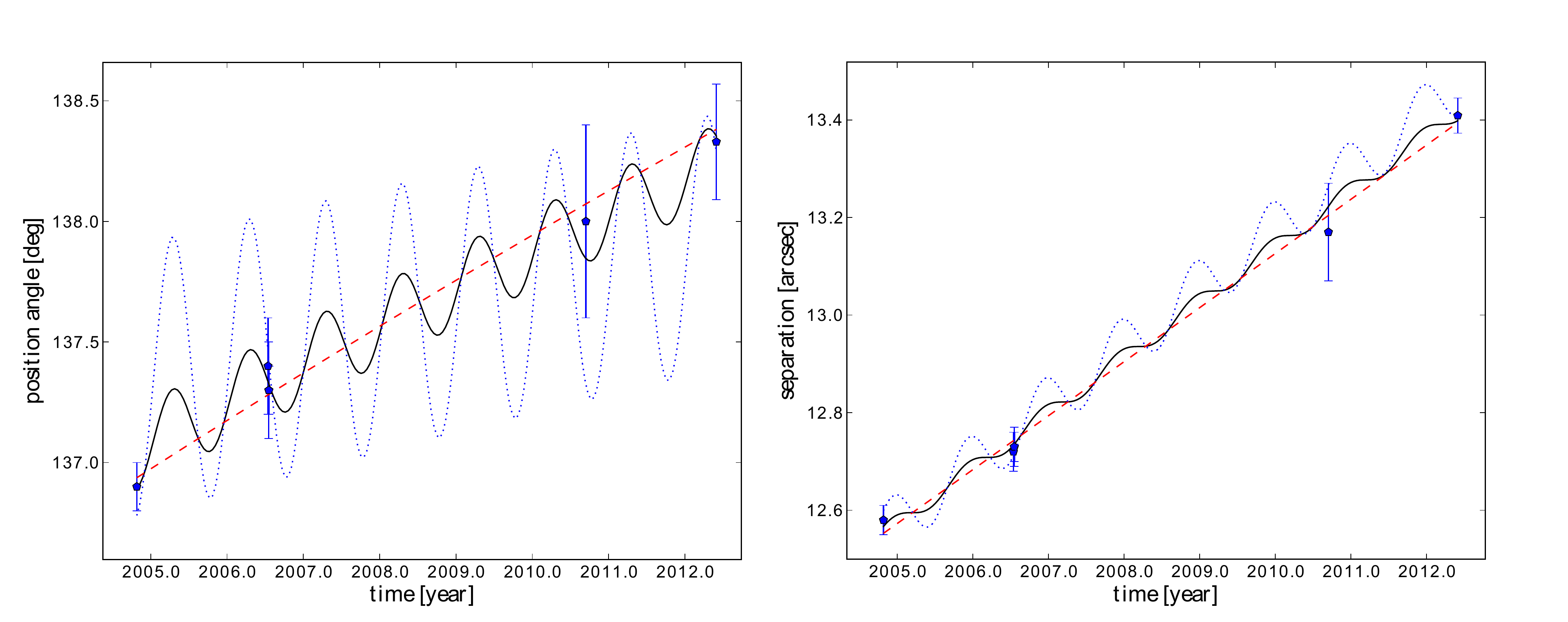}
\caption{{\bf Change in position angle (left) and separation (right) with time:}
The five astrometric data points for the object called Fomalhaut b are shown (epochs as listed in Table 1) 
in order to try to measure its own proper motion and parallax.
Differential proper motion and differential parallactic fits to the relative astrometric measurements of Fomalhaut A and b.
The solid (black) line shows our best fit of a differential proper motion of $-50.4$\,mas/yr in RA and $107.8$\,mas/yr in Dec 
as well as a {\em differential} parallax of $39.1$\,mas, yielding a total distance of 11 pc for Fomalhaut b.
For comparison, we also show a fits with no differential parallax (dashed red line) and with the maximum possible differential  
parallax of $129.8$\,mas (dotted blue line), 
corresponding to a distance estimate of 8 pc in the first case and any (larger) distance in the second case.
A differential parallax between Fom A and b that is equal to the measured absolute parallax of Fom A implies 
no measureable parallax of Fom b. Then, Fom b could be located at any (larger) distance.
In the first (more likely) case the differential motion in RA and declination is $-42.8$\,mas/year
and $110.9$\,mas/year, respectively. In the second case it would be $-68.1$\,mas/year and $100.5$\,mas/year, respectively..
The relative change in position of Fomalhaut A and b can be due to different distances.
The best fit shown results in $\sim 11$ pc as distance for the companion candidate, but it is not significant. 
Within $\ge 2~\sigma$ error bars, any other larger distance is also possible,
}
\end{figure*}

\subsubsection{X-ray data}

Young and nearby NSs are detectable as bright X-ray sources (e.g., Walter et al, 1996, Haberl et al. 1997, and Haberl 2007 for a review). 
Therefore, we checked the X-ray archives whether there is a source located at the position of Fomalhaut~b. 
Only one 1.5~ks {\it EINSTEIN} IPC (Miller et al. 1978) pointing from the year 1979 and a 6.2~ks PSPC exposure with 
ROSAT (Tr\"umper 1983) from January 1996 are available in the archive (in addition to a 170~sec exposure from the ROSAT All-Sky-Survey), 
see Figs. 2 and 3. The {\it EINSTEIN} IPC observation shows many artifacts that mimic sources, but there is no evidence of X-ray emission 
at the current or past position of Fomalhaut~b (Fig. 2). Two potential X-ray sources (denoted as ``source 1" and ``source 2", respectively) 
in Fig. 2 are too distant from Fomalhaut~b's past position, considering its proper motion, to be identified with Fomalhaut~b.

Also the ROSAT PSPC data give no evidence of X-ray emission at the position of Fomalhaut~b. 
Furthermore, ``source 1" (out of view) and ``source 2" detected in the {\it EINSTEIN} IPC pointing are also not visible, 
suggesting that the latter is an artifact or a variable source (Fig. 3).

\begin{figure*}
\centering
\includegraphics[width=0.96\textwidth, viewport=8 88 695 520]{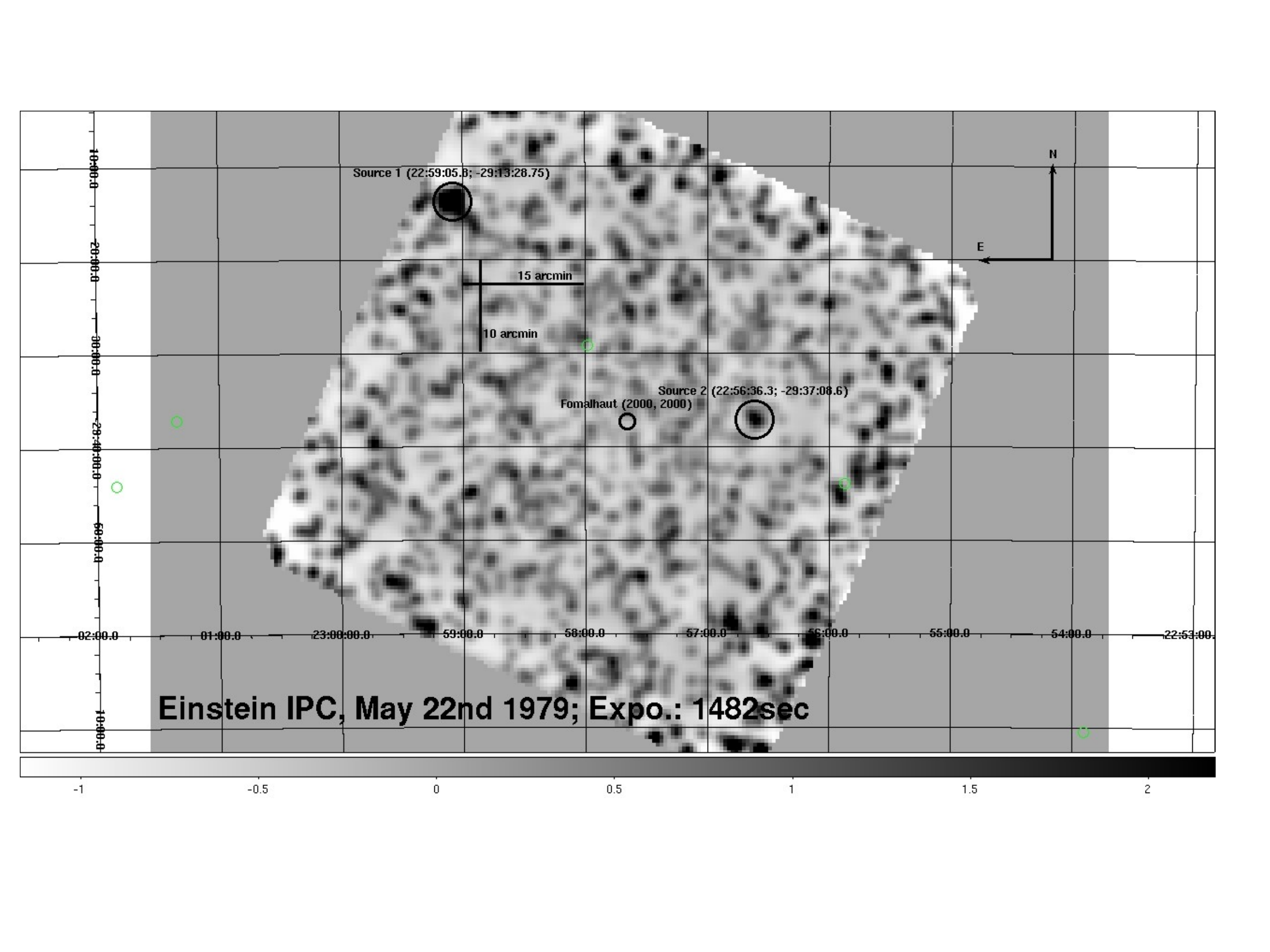}
\caption{{\bf X-ray observation with Einstein:} A 1.5~ks {\it EINSTEIN} IPC pointing exhibits numerous artifacts and two potential X-ray sources.
However, there is no evidence for X-ray emission at Fomalhaut's former position (1979) in the center of this image.}
\label{e}
\end{figure*}

\begin{figure*}
\centering
\includegraphics[width=0.75\textwidth]{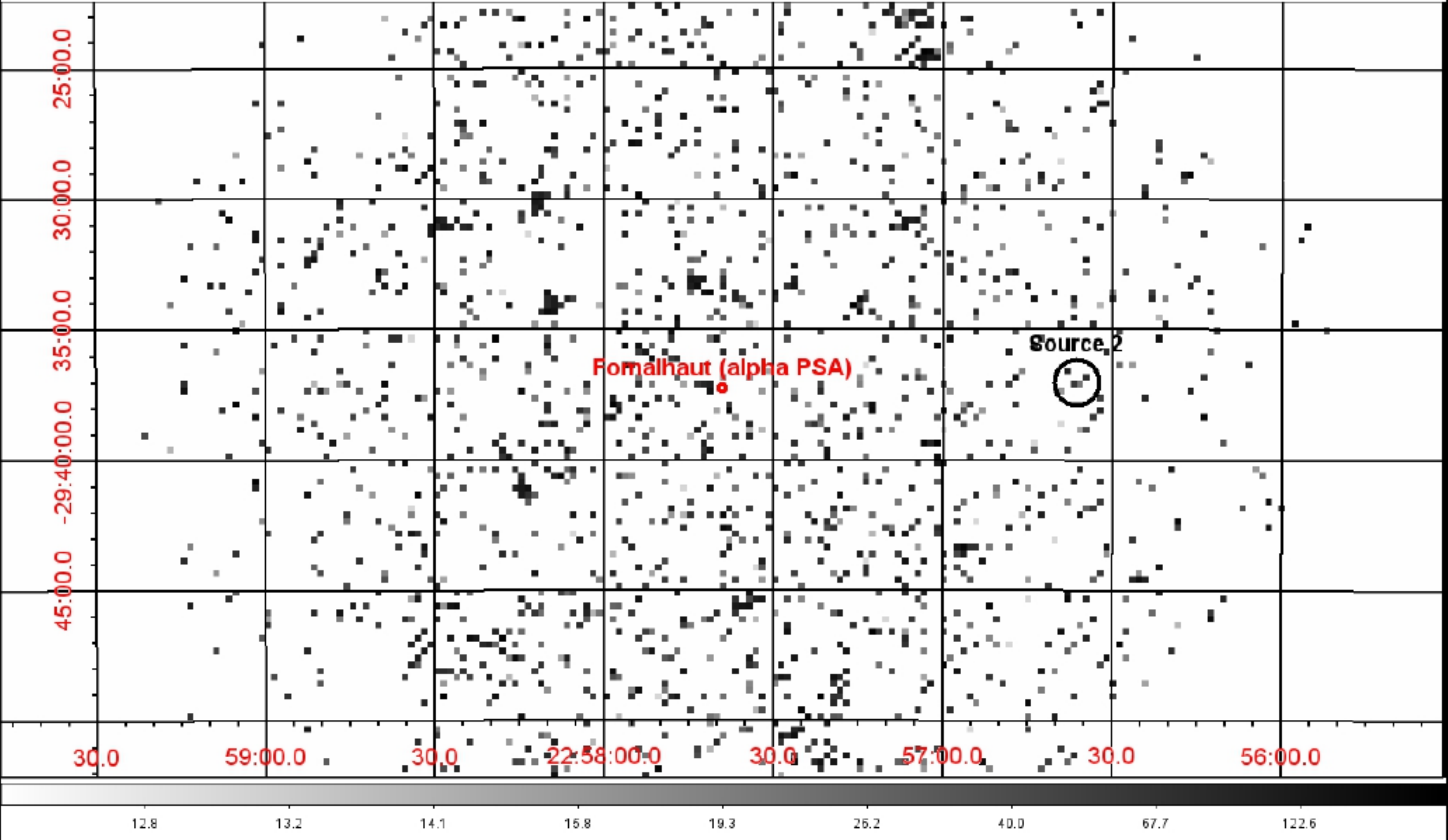}
\caption{{\bf X-ray observation with ROSAT:} 
Data from {\it ROSAT} PSPC (6283~sec, 0.11-2.4~keV, 1996 Jan 19) do not show X-ray emission near Fomalhaut ($\alpha$ PsA).
Fomalhaut b is 13 arc sec NW of A, there is neither a source nor more background photons.
``Source~1" from \autoref{e} is not in the field of view and ``source~2" is not detected even though the exposure time is larger.}
\label{r}
\end{figure*}

Based on these non-detections of Fomalhaut~b in the X-ray images, one can put rough constraints on the properties of the putative NS. 
In the 6.2~ks ROSAT exposure obtained with the Boron filter (which blocks about $90~\%$ of the soft flux below 0.3 keV), 
an upper limit count rate of $\le 0.00066$ cts/s was determined (Schmitt 1997); 
while this upper limit was determined for the star Formalhaut A, it should also apply to Fomalhaut b given the small separation
(see also Fig. 3, where no source is detected).

The isolated NS RXJ~1856.4-3754 appears as bright source with pure blackbody emission (therefore often serves as calibration target, see Mereghetti et al. 2012) 
in the X-ray energies with a ROSAT PSPC count rate of 3.67 cts/sec in the 0.11-2.4~keV band.
Hence, RXJ~1856.4-3754 is at least 556 times brighter than Fomalhaut~b in the ROSAT PSPC energy band (having taken into account that
RXJ~1856.4-3754 was observed without Boron filter, while Fomalhaut was observed with Boron filter). 
According to the most recent parallax measurements by Walter et al. (2010), RXJ~1856.4-3754 has a distance of $\sim 123$ pc, 
yielding an emitting area of $\sim 4.4$ km as origin of the X-ray radiation. 
Assuming that Fomalhaut~b emits as blackbody, too, 
the temperature $\mathrm{T_{\infty}}$ of its X-ray emitting area must be
below 380,000~K, if its radiation would have the same 
normalization\footnote{Due to high gravity and curved space around a NS, an observer at infinity
measures temperature $\mathrm{T_{\infty}=T\sqrt{1-r_s/R}}$ and radius $\mathrm{R_{\infty}=R/\sqrt{1-r_s/R}}$, 
where $\mathrm{r_s}$ is the Schwarzschild radius.} of $\mathrm{N=(4.4~km/123~pc)^2}$ as RXJ~1856.4-3754, 
or its luminosity must scale with $\mathrm{f<(4.4~km/123~pc)^2\times(10^{6}~K)^4}$ 
(Tr\"umper 2003, Tr\"umper et al. 2004, Walter et al. 2010) to obey the upper limit derived from the {\it ROSAT} data.

The non-detection of Fomalhaut b as NS is consistent with a NS with at least a few Myr 
age - even if only slightly background to Fomalhaut A, see below.
This also applies, if it would be a NS of a different kind, i.e. other than RXJ~1856.4-3754 and RXJ~0720.4-3125.
The latter two have kinematic ages of $0.5$ to 1 Myr (Table 4), while the X-ray non-detection
constraint applies to all NSs older than $\sim 10^{5.5}$ yr (Sect. 3.2.4).

The fact that Fomalhaut\,A (spectral type A4) is not detected in X-rays is not surprising, 
since A stars do not have strong hot winds nor a corona. 
However, some A4V stars exhibit X-ray luminosities that would correspond to about 50 times the detection limit 
of Fomalhaut\,A (Schr\"oder \& Schmitt 2007). Many 
of those (if not all) X-ray detected A-type stars are considered to host a very close stellar companion;
this is not the case for Fomalhaut\,A; there are, however, Fomalhaut B as a wide K4-type stellar companion (0.3 pc away, Mamajek 2012)
and Fomalhaut C (LP 876-10) as 2nd wide M4-type stellar companion (0.77 pc away, Mamajek et al. 2013).

\subsubsection{Photometry and SED}

Absorption/extinction caused by the ISM must be taken into account (note that according to L\"ohne et al. 2012a,b extinction caused by the disk around Fomalhaut is negligible). 
We use the average Galactic extinction curve provided by Fitzpatrick \& Massa (2007). They fitted this extinction curve with a spline function based on several 
anchor points that were calculated by comparing model spectra with measured data from individual reference stars. 
The fit errors are in the order of several mmag, whereas we stress that the extinction curves for individual stars 
(hence, individual directions) can deviate significantly (Fitzpatrick \& Massa 2007).

Taking blackbody normalization, temperature (of the optically emitting area), and A$_{\rm V}$ as free input parameters, 
the resulting magnitudes have to fit those measured by Currie et al. (2012) and Kalas et al. (2008).\footnote{We have tested 
and verified the fit procedure with the optical data of RXJ~1856.4-3754, its optical magnitudes, the distance 123 pc, 
a radius of 17 km, and negligible extinction.} 
As an estimate for the fit quality, we introduce the factor\footnote{The low number of data points prevent a fit quality estimate in terms 
of reduced $\mathrm{\chi^2}$.} $\mathrm{q^2=1/k^2\sum_{i}^{k}(m_i - \bar{m}_i)^2}$, where $\mathrm{m_i}$ are the 
measured magnitudes (Currie et al. 2012, Kalas et al. 2008) and $\mathrm{\bar{m}_i}$ the fitted magnitudes in the filter $i$. 
Since Fomalhaut~b is only detected in three filters, the total number of different magnitudes is $k=3$.
Distance, temperature (of the optically emitting area), and extinction are free fit parameters, the radius was assumed to be 17 km.

Currie et al. (2012) and Kalas et al. (2008) list magnitude errors in the order of 0.1~mag to 0.2~mag for the HST photometry (see Table 1). 
However, the magnitudes of the same filter differ by 0.5~mag for different measurements and by different authors, 
suggesting that the systematic errors are much larger than the statistical errors. 
We calibrated the modelled blackbody flux to the Vega magnitude system (our results were checked with the average Vega flux densities in the different 
HST filters as listed in the HST handbook), since Currie et al. (2012) and Kalas et al. (2008) give Fomalhaut~b's magnitudes in the Vega system, 
and corrected for non-infinite aperture (see HST handbook), see Table 1. 
Furthermore, we calculated the HST magnitudes of the two optically detected isolated NSs with known distance, see Table 2, 
and compared our results to those given by Kaplan et al. (2011) in ST the magnitude system as an additional check.

We can find blackbody spectral energy distribution which fit the observed data,
they are in the parameter range of $\mathrm{T_{\infty}=6,810-126,500~K}$, $\mathrm{D=1.6-33~pc}$ and $\mathrm{A_V<2.5~mag}$ 
(but not all combinations in the parameter ranges are possible). 
In Table 3, we show the ten allowed combinations as examples. 
The resulting 
effective temperatures (of the optical emitting area)
yield ages of at least 
$\sim 10^{5.5}$ yr
for the putative NS 
according to cooling curves in Aguilera et al. (2008) and Page et al. (2009); 
however, since NSs cool very rapidly for 
effective temperatures (of the optical and/or X-ray emitting area)
of $\le 100,000$ K and ages 
above at around that age, 
a precise age estimate from the temperature is not well possible.
X-ray non-detection would not be surprising for Fomalhaut b being a relatively old NS.
We would like to point out that we assume just one (effective) temperature for both the
optical and X-ray emitting area, i.e. the whole neutron star surface.
This temperature fits the optical magnitudes known and agrees with the X-ray upper limit as observed.
It is of course possible that the polar caps are hotter than the remaining surface area, 
so that the object as neutron star would show pulses, even if not yet detected.

A more realistic model of the emission of a young NS is a two component model: 
a cool ($\mathrm{T_{\infty}\approx300,000-400,000~K}$) blackbody emitting in the optical (representing the major part of the NS surface) and 
a hot ($\mathrm{T_{\infty}\approx10^5-10^6~K}$) blackbody, visible at X-ray energies, caused by 
the hot spot(s) at the magnetic poles (Kaplan et al. 2011). 
However, this would mean fitting three magnitudes with five parameters ($A_V$,$D$ or $\mathrm{R_{1\infty}}$,$\mathrm{R_{2\infty}}$,$\mathrm{T_{1\infty}}$,$\mathrm{T_{2\infty}}$) 
and -- as expected -- did not lead to useful results: 
E.g. any high effective temperature value 
(of the optical and X-ray emitting area, i.e. the whole surface)
can be compensated by large $\mathrm{A_V}$ values flattening the SED. 
To put constraints on the two blackbody model it is necessary to derive more data points for the SED, in particular at UV energies.

\begin{table}
\centering
\caption[]{Ten allowed parameter combinations consistent with the photometry of Fomalhaut b.
We list magnitudes, effective temperature ($\mathrm{T_{\infty}}$) 
(of the optical emitting area)
as seen from an observer at infinity, 
predicted distance to the Sun ($D$), and interstellar extinction  $\mathrm{A_V}$.}
\label{best_ten}
\begin{tabular}{ccccccc}
\hline
F435W & F606W & F814W & $\mathrm{T_{\infty}}$ & $D$ & $\mathrm{A_V}$ & $q$	\\
{[mag]} &	[mag] & [mag] & [K]	 									 & [pc]& [mag] 					& \\  
\hline
25.13 &	25.07 &	24.87 &	19360 &	6.2 &	0.53 &	0.045\\
25.15 &	25.09 &	24.87 &	37890 &	9.0 &	0.85 &	0.048\\
25.12 &	25.02 &	24.81 &	14320 &	4.9 &	0.33 &	0.050\\
25.12 &	25.08 &	24.87 &	45710 &	10.1 &	0.86 &	0.051\\
25.22 &	25.11 &	24.84 &	54190 &	10.5 &	1.04 &	0.052\\
25.24 &	25.12 &	24.87 &	23840 &	6.8 &	0.77 &	0.053\\
25.07 &	25.02 &	24.86 &	13960 &	5.1 &	0.22 &	0.054\\
25.20 &	25.13 &	24.89 &	55520 &	11.1 &	0.97 &	0.054\\
25.09 &	25.09 &	24.92 &	62010 &	12.4 &	0.85 &	0.058\\
25.11 &	25.03 &	24.79 &	53960 &	10.4 &	0.98 &	0.059\\
\hline
\end{tabular}
\end{table}

\begin{figure*}
\centering
\includegraphics[angle=270,width=0.96\textwidth]{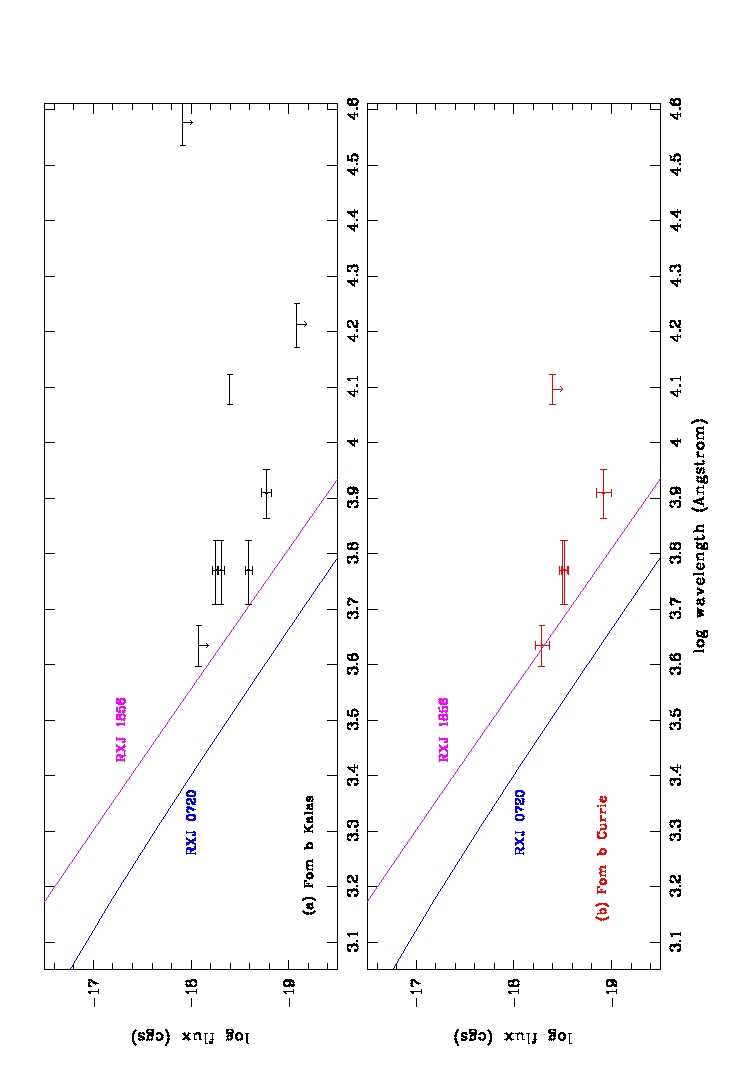}
\caption{{\bf Optical photometry for the object called Fomalhaut b compared to two neutron stars:}
(a, top) in black (Kalas et al. 2008) and
(b, bottom) in red (Currie et al. 2012)
with IR upper limits as arrows (band width indicated), data from Table 1.
The optical emission from RXJ~1856.4-3754 can well be fitted with a 380,000 K blackbody, the emission from RXJ~0720.4-3125
is consistent with 112,000 K (both assumed to be unabsorbed); we show their spectral energy distribution 
(Planck functions for 112,000 K and 125,500 K) as pink and blue line, respectively. See also Fig. 5.}
\end{figure*}

\begin{figure*}
\centering
\includegraphics[angle=270,width=0.96\textwidth]{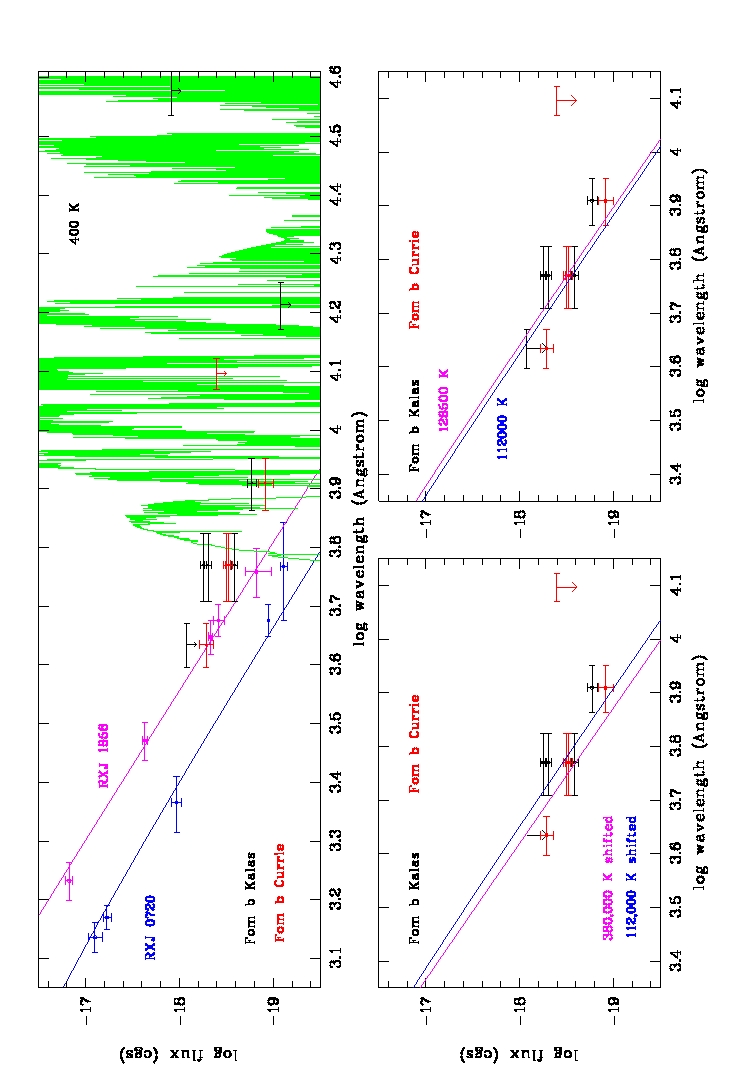}
\caption{{\bf Fomalhaut b photometry compared to typical hot neutron stars and a typical cold planet.}
The available optical photometry for the object called Fomalhaut b in red (Currie et al. 2012) and black (Kalas et al. 2008)
with IR upper limits as arrows (band width indicated), data from Table 1, as also plotted in Fig. 4.
We also show the optical data points for the two well-known isolated middle-aged neutron stars
RXJ~1856.4-3754 and RXJ~0720.4-3125 in pink and blue, respectively (data from Table 2).
The optical emission from RXJ~1856.4-3754 can well be fitted with a 380,000 K blackbody, the emission from RXJ~0720.4-3125 
is consistent with 112,000 K (both assumed to be unabsorbed).
The model atmosphere for 400 K in green (here from AMES COND for log g = 4.0 (Chabrier et al. 2000, Allard et al. 2001) for a few hundred Myr planet,
from phoenix.ens-lyon.fr/Grids/AMES-Cond/SPECTRA, as an example (similar for other models), 
scaled in y-axis as far down as possible to still fit the optical data points) 
does not fit the Fomalhaut b data as known before -- it is inconsistent with the IR non-detection.
We show all data in the upper panel and then again (for clarity) in the lower panels the Fomalhaut b data 
with Planck functions for 380,000 K and 112,000 K (bottom left) and with Planck functions for 112,000 K and 125,500 K (bottom right),
which do fit 
(fits with 17 km radius as RXJ~1856.4-3754). 
The y-axis (flux) ranges are identical, the x-axis (wavelength) ranges differ slightly.}
\end{figure*}

\subsubsection{Constraining the neutron star properties}

Given the observed astrometry and photometry, we can now try to constrain the properties of Fomalhaut b, 
if it would be a NS, in particular its distance and age range. 
We assume for most part of this section that Fomalhaut b as NS would have the typical radius of NSs, $\sim 10$ to 17 km.
First, we assumed $\sim 17$ km radius for the optically emitting area and $\sim 4.4$ km as radius of the 
X-ray emitting area in our spectral fits; these are values similar as for RXJ~1856.4-3754 (and maybe RXJ~0720.4-3125).
Afterwards, we compare the object called Fomalhaut b with other neutron stars,
which have smaller and/or hotter polar caps than RXJ~1856.4-3754 and RXJ~0720.4-3125.
At the very end of this subsection, we also consider even smaller emitting areas (and smaller radii)
like in strange (quark) stars.

The non-detection of a supernova remnant places a lower limit to the age of a NS to roughly $10^{5}$ yr, 
a supernova remnant has diffused and faded away. In such a case, one would regard this NS (Fomalhaut b) as middle-aged,
isolated NS ({\em isolated} means that there is neither a companion nor a supernova remnant).
Given the small distance of Fomalhaut b (even as NS), $\sim 11$ pc being our best fit, see above, the supernova remnant
would have a large extend on sky: The Vela remnant at a distance of $\sim 290$ pc (Caraveo et al. 2001, Dodson et al. 2003)
and an age of $\sim 11$ kyr (Dodson et al. 2002), both measured for the Vela pulsar, has an apparent size of $255^{\prime}$ (Green 2009).
A supernova remnant some $\sim 26$ times closer would have a size of $\sim 111^{\circ}$.
Even at such a large size, it might have been noticed in the ROSAT All-Sky Survey, but no such (large) remnant was detected
(in particular not at that position), see e.g. Busser (1998) and Schaudel et al. (2002).

If such a large remnant would not have been detectable, Fomalhaut b as NS would still not be a young NS,
because it would then be bright in X-rays (if young), which is not the case.
Hence, if a NS, Fomalhaut b is most likely middle-aged or old.

If Fomalhaut b as NS would be related to the Local Bubble ($\sim 50$ Myr old), a volume around the Sun with very low interstellar
medium density, then it might be possible that its SN did not form a detectable remnant; in this case, the NS might be younger than 50 Myr
(but still older than $\sim 10^{5.5}$ yr due to X-ray non-detection).

The non-detection of radio pulsations 
may simply be due to the fact that the pulses
are not beamed towards Earth; otherwise, they place a lower limit to the age of a NS to roughly $10^{8}$ yr, 
the so-called death-line or graveyard of radio pulsars:
Most known pulsating NSs are younger than $10^{8}$ yr given their (characteristic) spin-down age (Gyr old milli-second pulsars recycled
by mass transfer from their (former) companion are exceptions). 
In such a case, one would regard this NS (Fomalhaut b) as middle-aged to old.

The proper motion is definitely in the possible range for NSs. 
The astrometry (parallactic motion) would be well consistent with 
$\sim 11$ pc (best fit),
but also much larger distances are not excluded (Fig. 1).

In principle, the color of Fomalhaut b could also be compared to the NSs RXJ~1856.4-3754 and RXJ~0720.4-3125;
however, Kalas et al. (2008) and Currie et al. (2012) do not agree well on the magnitudes
and the differences and error bars in their values are on the order or larger than the colors,
and also, while Fomalhaut b was detected by HST with F606W, F435W, and F814W, neither RXJ~1856.4-3754 nor RXJ~0720.4-3125
were observed with F435W nor F814W (but detected in F606W).
We have re-reduced the HST photometry and arrived at values close to those of Kalas et al. (2008) and Currie et al. (2012),
but also our photometry error bars are comparable to the error bars and absolute differences between the results in
Kalas et al. (2008) and Currie et al. (2012).

The optical detections and the upper limits in the infrared and X-rays allow good fits for 
a range in effective temperature 
(of the optical and X-ray emitting area)
of up to roughly $\sim 100,000$ K with small to negligible extinction (see Table 3 for a few examples);
this is compatible with the more conservative X-ray temperature upper limit (up to 380,000 K)
For a NS, this would then yield a distance of up to 33 pc and an age of 
at least $\sim 10^{5.5}$ yr
(according to the cooling
curves in Aguilera et al. 2008 and Page et al. 2009), NSs start to cool very rapidly for temperatures $\le 100,000$ K
(of the optical and X-ray emitting area)
at ages somewhere between $10^{5.5}$ to $10^{6.5}$ yr, so that a precise age estimate from the temperature 
(of the optical and X-ray emitting area)
is hardly possible in this regime.

The non-detection in X-rays (Figs. 2 and 3) can place limits on temperature 
(of the X-ray emitting area)
and distance by 
comparison with other middle-aged isolated NSs (all without supernova remnant),
namely the NSs RXJ~1856.4-3754 and RXJ~0720.4-3125, which are middle-aged, isolated, and which have a known distance;
in Table 4, we list their ROSAT PSPC count rates, distances, ages, and V-band magnitudes -
to be used to scale to Fomalhaut b (we assume negligible extinction for Fomalhaut b and the NSs here).

Given that RXJ~1856.4-3754 and Fomalhaut b have very similar optical photometric magnitudes (Tables 1 and 2),
we can relate distances $d$ and temperatures $T$ (since flux scales with $T^{4}$ and $d^{-2}$).
If the temperature ratio between RXJ~1856.4-3754 (380,000 K) and Fomalhaut b (say 100,000 K to 150,000 K) is the same for the warm surface
responsible for the optical emission as for the hot polar spots responsible for X-ray emission, then we can scale
from the temperature ratio and the distance ratio (123 pc for RXJ~1856.4-3754 and, say, 11 pc for Fomalhaut b as NS)
as well as the X-ray count rate of RXJ~1856.4-3754 (3.67 cts/sec, Table 4)
also to the expected X-ray count rate of Fomalhaut: With the PIMMS software,
we obtain $\sim 0.00066$ cts/sec for ROSAT PSPC with Boron filter for 112,000 to 126,500 K at $\sim 11$ pc;
this is exactly the upper limit count rate obtained for Fomalhaut (A and b): 0.00066 cts/sec (Schmitt 1997). 
Hence, for a distance range of 11 pc (best fit obtained from the astrometry),
a NS would need to have a temperature 
(of the X-ray emitting area)
of 112,000 K to 126,500 K to obey the X-ray upper limit.
The X-ray non-detection of Fomalhaut b is then consistent with being a NS.
Indeed, the temperature of 380,000 K is both the upper limit on the temperature of the X-ray emitting region
(based on the comparison with RXJ~1856.4-3754, above), and it is also close to the temperature of the
optical emitting region of RXJ~1856.4-3754, see e.g. Kaplan et al. (2011).

In Figs. 4 \& 5, we show the available photometry and upper limits of the object known as Fomalhaut b,
compared to the NSs RXJ~1856.4-3754 (380,000 K) and RXJ~0720.4-3125 (112,000 K) as well as compared to several blackbodies
with temperatures of 112,000 K to 126,500 K
(for the X-ray emitting area),
as well as a typical model atmosphere for 400 K (as should be expected for a planet, Kalas et al. 2008).
The blackbodies of 380,000 K to 112,000 K do fit the Fomalhaut b data.
For the comparison with RXJ~0720.4-3125, one should keep in mind
that Kaplan et al. (2011) showed that a Rayleigh-Jeans tail with a temperature of 112,000 K
would not fit the spectrum without an additional power law component.
For RXJ~1856.4-3754, however, there is no evidence for a deviation from a blackbody.

The constraints from optical data and X-ray non-detection are also consistent with an age above
$\sim 10^{5}$ yr as derived from the non-detection of a supernova remnant
or even
$\sim 10^{8}$ yr as derived from the non-detection of radio pulsations (if beaming towards us). 
By comparison with RXJ~1856.4-3754 and RXJ~0720.4-3125 (i.e. same temperature and area of the emitting polar caps),
Fomalhaut b as NS would be $\sim 10^{5.5}$ yr (or older) as derived from the non-detection of X-rays, but see below.

Let us now also compare the object called Fomalhaut b with NSs other than RXJ~1856.4-3754 and RXJ~0720.4-3125,
namely with NSs with smaller and/or hotter emitting areas (polar caps).
E.g., the radio pulsars PSR J0108-1431, PSR B1929+10 and PSR B0950+08 are detected in X-rays at large distance;
the existence of radio-silent NS with hot and/or small emitting regions is possible.
According to recent X-ray observations with Chandra and XMM, the relevant parameters are known.

PSR J0108-1431 can be fitted with a blackbody with $k \cdot T = 0.28$ keV and an X-ray emitting area of $53^{+32}_{-21}$ m$^{2}$,
or a power law with $\gamma = 2$ (Pavlov et al. 2009), it has an age of $\sim 160$ Myr and a distance of $\sim 210$ pc (Taylor \& Cordes 1993).
Then, using the PIMMS software, we expect $0.045$ count per second (0.027 to 0.072 for full $1~\sigma$ error range) with ROSAT PSPC
with boron filter (same setup as used in the observation of Fomalhaut), if such a NS would be at $\sim 11$ oc distance only. 
This is more than the ROSAT PSPC Fomalhaut upper limit being 0.00066 cts/sec (Schmitt 1997).
According to Posselt et al. (2012), this NS has an energy of $k \cdot T = 0.11$ keV with the radius of the X-ray emitting area being 43 m.
Then, we would obtain with PIMMS a ROSAT PSPC coutn rate of 0.0047 cts/sec with boron filter, again at 11 pc, again larger than the upper limit.
A NS like PSR J0108-1431 would have been detected up to 0.00066 cts/sec up to a distance of 90 to 120 pc,
but is not detected at the position of what is called Fomalhaut b.

PSR B1929+10 is $\sim 3$ Myr old at $\sim 361$ pc distance with hot polar caps ($\sim 0.3$ keV or $\sim 3.5 \cdot 10^{6}$ K, with
a projected emitting area of $\sim 3000$ m for the two X-ray emitting polar caps together, 
i.e. a radius of of the X-ray emitting area of $\sim 21.5$ m each for two circular polar caps),
resulting in an X-ray luminosity (or the polar caps) of $\sim 1.7 \cdot 10^{30}$ erg/s in the 0.3-10 keV band,
see e.g. Misanovic et al. (2008) or Slowikowska et al. (2005) as well as references therein.
Again using PIMMS, we would expect a ROSAT PSPC count rate of 3.47 cts/sec with the boron filter at 11 pc,
so that a NS like PSR B1929+10 would be detectable until $\sim 800$ pc at 0.00666 cts/sec (but is not detected at the Fom b position).

PSR B0950+08 is $\sim 17$ Myr old at $\sim 262$ pc distance with hot polar caps ($k \cdot T = 0.086$ keV or $\sim 10^{6}$ K 
with $\sim 250$ m radius of the X-ray emitting caps),
resulting in an X-ray luminosity of $\sim 3 \cdot 10^{29}$ erg/s from the polar caps,
see e.g. Zavlin \& Pavlov (2004) or Becker et al. (2004) as well as references therein. 
Again using PIMMS, we would expect a ROSAT PSPC count rate of 1.097 cts/sec with the boron filter at 11 pc,
so that a NS like PSR B1929+10 would be detectable until $\sim 450$ pc at 0.00666 cts/sec (but is not detected at the Fom b position).

By comparison with RXJ~1856.4-3754 (X-ray luminosity of $3.8 \cdot 10^{31}$ erg/s at 120 pc, Burwitz et al. 2003),
PSR B1929+10 would be detectable at up to five times larger distances than RXJ~1856.4-3754,
and PSR B0950+08 would be detectable at up to eleven times larger distances than RXJ~1856.4-3754.
PSR J0108-1431, PSR B1929+10, and PSR B0950+08 would have been detectable in X-rays at a distance of only $\sim 11$ pc
(best fit for Fomalhaut b) -- even given the short exposure time of the ROSAT pointing.
The astrometry of Fomalhaut b does not exclude larger distances, where those two NSs would also have been detectable.

Hence, if Fomalhaut b as NS would be similar to PSR J0108-1431, PSR B1929+10, or PSR B0950+08 
in both temperature and radius of both the X-ray emitting area
as well as age, than it would be detectable at 11 pc. Or, putting it another way around,
Fomalhaut b as NS would need to be older than the previously given lower limit of $\sim 10^{5.5}$ yr.
On the other hand, the available optical photometry of Fomalhaut b allows good fits only for blackbody 
temperatures (of the optically emitting areas) from 112,000 to 126,500 K, which may be too small for PSR B1929+10 and PSR B0950+08.

If Fomalhaut A is a member of the several hundred Myr ($440 \pm 40$ Myr) old Castor Moving Group (Barrado y Navascues 1998),
it might well be possible that Fomalhaut b as NS (and/or its progenitor) also belongs to this Moving Group -
given that Fomalhaut b (even as NS) and Fomalhaut A have a similar proper motion (as most members of
Moving Group have a similar proper motion). \\
The Castor Moving Group has 26 known members (plus the two stellar companions
to Fomalhaut A, being Fomalhaut B and C with K4 and M6) with at least eight A-type stars,
four stars with A0-2 (Barrado y Navascues 1998), so that it is not impossible that there was originally also one
early B-type star in this group (the progenitor of Fomalhaut b as NS):
We have converted the known spectral types of the Castor Moving Group members (Barrado y Navascues 1998, Mamajek 2012,
Mamajek et al. 2013) to main-sequence masses to investigate the present mass function; extrapolating
from the bin with the largest masses (a bin with stars above 1~M$_{\odot}$) by using the exponent $N \simeq 2.7 \pm 0.7$
from the Kroupa et al. (1993) initial mass function to even larger masses, we can estimate the expectation number
for core-collapse supernova progenitors (with at least 8~M$_{\odot}$) in the Castor Moving Group to be $0.2 ^{+0.6} _{-0.4}$
within $2\sigma$ error bars. Hence, it may appear unlikely, but possible. \\
If the progenitor of Fomalhaut b as NS would indeed have been a member of the Castor Moving Group, then Fomalhaut b as NS would now
have an age only slightly below the age of the Castor Moving Group (given the short life-time of its
progenitor), so that it would have an age of a few hundred Myr. Such an age is fully consistent with the cooling curves
of NSs given its brightness in the optical and its non-detection in the X-rays.
(The proper motion of the two presumable stellar companions Fomalhaut B and C being co-moving with Fomalhaut A
can be interpreted either that they form a triple stellar system (with undetected orbital motion around the
common center-of-mass) or that all three stars are (independant) members of the same Moving Group.)
The proper motion of Fomalhaut A and b would be equivalent to a tangential (2D) velocity of $\sim 30$~km/s 
for $\sim 11$ pc distance,
which is a low, but a possible velocity for a NS. The unknown radial velocity has to be added.
Only if Fomalhaut b (as NS) got a very small kick in its supernova, then its velocity can now still be similar
to the velocity of the progenitor star, i.e. the typical Castor velocity; a small velocity would be consistent
with a small SN kick. \\
We would like to stress that, for our interpretation of Fomalhaut b as NS, it is not essential that it would
be a member of the Castor Moving Group. \\
At an age of $440 \pm 40$ Myr (or, say, hundreds of Myr), detections of radioisotopes on Earth 
due to the very nearby SN explosion is also very difficult:
At the proper motion and current distance of the Fomalhaut companion candidate,
it would have moved $\sim 16$ kpc in 440 Myr
(the space velocity used considers only the known two-dimensional motion, not its unknown radial velocity); 
it would have orbited the Galactic Center almost twice, 
so that it is not possible to constrain well the location of the SN; hence, it is also completely unknown,
whether the SN took place within a few tens or hundreds of pc around Earth. Also, Firestone (2014) had to restrict their study 
to SNe (in radioisotopes) within the last 300 kyr given also the half lifes and measurement precision of relevant radionucleids.

Could Fomalhaut b as NS be just $\sim 2$ Myr young~? 
It is at least $\sim 10$ times closer than RXJ~0720.4-3125 (at $\sim 280$ pc, Eisenbeiss 2010),
so that it would be expected to be at least $\sim 100$ times brighter, if at the same age ($\sim 1$ Myr, Table 2)
and with the same temperature 
of the emitting area
(Figs. 4 \& 5), and radius ($\sim 17$ km, Eisenbeiss 2010).
In fact, it is only $\sim 3$ times brighter in the optical than RXJ~0720.4-3125 ($\sim 1$ Myr, Tables 1, 2, and 4).
If Fomalhaut b as NS would be 2 to 3 times smaller than RXJ~0720.4-3125, it would be 4 to 9 times fainter. 
The remaining factor could easily be obtained by a slightly different temperature 
of the emitting area
(luminosity scaling with
the forth power of the temperature), even if at the same radius; also, cooling tracks at that age are
highly uncertain. Hence, from these considerations, Fomalhaut b as NS could be as young as $\sim 2$ Myr. \\
If Fomalhaut b as NS would be as young as $\sim 2$ Myr, it would have moved $\sim 61$ pc since 2 Myr
(the space velocity used considers only the known two-dimensional motion, not its unknown radial velocity).
Given its current distance of $\sim 11$ pc (best fit), 
it had a distance of somewhere with $\sim 72$ pc at birth (i.e. still inside the Local Bubble).
A supernova that close might have left some effect on Earth, e.g. a $^{60}$Fe signal,
such as the one detected at an age of 2 Myr under the Earth ocean crust (Knie et al. 1999, Fields 2004, Bishop et al. 2013).
We conclude that, for certain parameter combinations,
Fomalhaut b as NS could be 2 Myr young and could be the NS that was born 
in the nearby supernova that left $^{60}$Fe on Earth. However, this is speculative.
The true age could be constrained with, e.g., an X-ray detection.

If Fomalhaut b as compact, non-planetary object would not have the typical NS radius, nor the same radius as
the NSs RXJ~1856.4-3754 or RXJ~0720.4-3125, to which we compared it, but a different radius, also a different
distance estimate would apply. Theoretical considerations show that a NS cannot be much larger than 
RXJ~1856.4-3754 ($17 \pm 3$ km, Tr\"umper et al. 2004, Walter et al. 2010). However, some equations-of-state
predict smaller radii such as half the radius, namely for strange (quark) stars.
Our considerations above hold in a similar way for both normal NSs and strange stars, with one exception:
If strange stars are half a large as normal NSs, then the allowed distance range for Fomalhaut b as strange star
would need to be accordingly smaller by a factor of $\sqrt 2$.

For similar temperatures 
(of both the X-ray and optical emitting areas),
the radius of the emitting area cannot be smaller by more than a factor of 5 than in RXJ~1856.4-3754: 
A somewhat larger distance of, say, 20 pc -- devided by the square root of 5 -- gives 8 pc, the lowest allowed distance
(at or just behing Fomalhaut A).

If Fomalhaut b would be a NS or a strange star, we would not expect photometric variability (except maybe pulsations),
and we would not expect the object to be resolved/extended.

\begin{table*}
\caption{Properties of two isolated neutron stars. We list both the kinematic age from tracing back the motion of the object to its presumable birth place
inside an OB association as well as its characteristic (spin-down) age $\tau _{ch}$, which is to be considered an
upper limit to the true age; the kinematic ages fit better with cooling curves than spin-down ages (Tetzlaff et al. 2010). \\
References: Wal96: Walter et al. 1996, Wal10: Walter et al. 2010, vKK08: van Kerkwijk \& Kaplan 2008, vKK01;
Tet10: Tetzlaff et al. 2010; van Kerkwijk \& Kulkarni 2001, Hab97: Haberl et al. 1997, Eis10: Eisenbeiss 2010,
Kap05: Kaplan \& van Kerkwijk 2005; Tet11: Tetzlaff et al. 2011.}
\begin{tabular}{l|cc|cc|cccc|cc} \\ \hline
NS      & X-ray   & Ref  &   distance              & Ref   & \multicolumn{4}{c}{age [Myr]}       & F606W & Ref  \\ 
name    & cts/sec &      &   [pc]                  &       & $\tau _{ch}$ & Ref & kin. & Ref         & [mag] &      \\ \hline
RXJ1856 & 3.67    & Wal96 &   $123 ^{+15} _{-11}$  & Wal10 & 3.8        & vKK08 & $\sim 0.5$ & Tet10 & 25.6  & vKK01 \\
RXJ0720 & 1.65    & Hab97 &   $280 ^{+210} _{-85}$ & Eis10 & 1.9        & Kap05 & 0.7-1.0    & Tet11 & 26.8  & Eis10 \\ \hline
\end{tabular}
\end{table*}

\subsubsection{Probability considerations}
From the age of the Galaxy and the core-collapse supernova rate of a few events per century (Tammann et al. 1994),
as well as from several other considerations (metallicity of the ISM, pulsar birth rates, $^{26}$Al content, etc.),
$\sim 10^{8}$ NSs should have been produced in our Galaxy so far. 
Of course, only the hottest (i.e. youngest) ones can be detected in the optical:
For the bright isolated NS RXJ~1856.4-3754, the optical emission and size indicates a surface temperature 
(of the optical emitting area)
of a few $10^{5}$ K 
(Tr\"umper 2003, Tr\"umper et al. 2004, Walter et al. 2010).
According to several sets of NS cooling curves, and depending on assumption about their interior and their mass 
(see, e.g., cooling curves in Aguilera et al. 2008 and Page et al. 2009),
NSs with that temperature 
(of the optical emitting area)
might be sufficiently hot and, hence, detectable in the optical (like RXJ~1856.4-3754), until an age of $\sim 10^{6}$ yr
(or a few times $\sim 10^{6}$ yr).
For a constant NS formation rate since $\sim 1.4 \cdot 10^{9}$ yr (the age of the Galaxy), 
there should then be some $\sim 72,000$ detectable (young, hot) NS.
Given the velovity distribution of NSs, many of them can leave the Galaxy,
but the young ones that we consider here could not leave it, yet.
Given their high velocities, NSs are not restricted to the Galactic plane; 
if they would be distributed uniformly on the sky ($4\pi$~sr = 41253 square degree), 
then we would expect $\sim 1.75$ detectable NS per square degree.

To get a rough estimate of the area that was covered by instruments capable of detecting Fomalhaut\,b, 
we used the HST archived exposure catalog (provided by STScI in 2007). 
This is a sensible approach, since Fomalhaut\,b has so far only been detected with the HST, only in the optical,
and the majority of the HST measurements of Fomalhaut\,b including the discovery epoch were taken before 2007.
From the catalog, we extracted all imaging exposures which were taken in bands with a central wavelength shorter than 1\,$\mu$m 
and with longer exposure times than 1000\,s, i.e. exposures in which objects as faint as Fomalhaut\,b should have been detected. 
After removal of duplicate exposures (i.e. exposures with a separation of less than 1\,arcmin), we found 862 exposures with ACS/HRC, 
2216 exposures with ACS/WFC, and 2900 exposures with WFPC2 matching our criteria. 
Given the fields of view of these instruments, these exposures cover a total area of 11.65 square degrees on the sky. 
Thus, $\sim 20$ NSs could be contained in this area. 
A few well-known middle-aged NSs are indeed optically detected by HST (like RXJ~1856.4-3754
and the other the so-called Magnificent Seven NSs, see review in Haberl 2007), namely in deep exposures,
some of them at high galactic latitude.

If we restrict this estimate to the Galactic plane ($|b| \le 20^{\circ}$), where almost all stars (and most NSs) are located (we deal here with a 
potential NS in the background to a star with small apparent separation to that star), then we are left with only 1.64 square degrees,
so that we would expect $\sim 3$ detectable NSs.
Therefore, it is not unlikely to discover a previously unknown NS in one or a few of these exposures.

If we would restrict the estimate to a small area on the sky around a bright nearby star (like Formalhaut A), we can arrive at a different estimate:
Fomalhaut b is some 13 arc sec off Fomalhaut A, so that a circle with 13 arc sec radius is relevant here for a background probability estimate.
The probablity to find any one of $\sim 72,000$ detectable NSs (as given above) within 13 arc sec around one particular star 
(on the whole sky) is then only $\sim 7 \times 10^{-3}$.
However, several thousand different young and/or nearby stars were surveyed with deep exposures for planetary companions by direct imaging,
e.g. the directly imaged planet candidate (or brown dwarf) companion near GQ Lup was first detected in 1999 by the HST in an optical imaging snapshot program
(Neuh\"auser et al. 2005).
Then, the expectation number of background (or foreground) NSs within 13 arc sec around any one of several thousand stars (say, 5000 stars)
is $\sim 0.35$. This estimate is not significantly different from one object found.

Hence, probability estimates show that it is not very unlikely to find one unrelated object (or even a NS)
with deep imaging in the background (or foreground) of one of the surveyed stars.

Above, we have estimated the probability for young, self-luminous NSs only. It is also possible that older NSs get
re-heated due to Bondi-Hoyle accretion of interstellar material (e.g. Madau \& Blaes 1993).
However, it may well be that re-heating of old NSs is not important, because otherwise many more such
re-heated NSs should have been detected by, e.g., the ROSAT All-Sky Survey and other X-ray missions
(see e.g. Neuh\"auser \& Tr\"umper 1999) -- such old accreting NSs were not detected (in particular not in large numbers)
probably due to both higher space velocities $v$ (Bondi-Hoyle accretion scales with $v^{-3}$)
and larger magnetic fields (propeller effect) than assumed in Madau \& Blaes (1993).
On the other hand, with PSR J0108-1431 at $\sim 160$ Myr, there also exists an example of a
relatively old NS which is still quite hot -- probably due to internal re-heating
(Tauris et al. 1994, Pavlov et al. 2009, Posselt et al. 2012). 
If we would add those NSs in our probability estimate, the probability for finding one NS would increase.

\begin{figure}
\centering
\includegraphics[width=0.48\textwidth, viewport=90 265 480 560]{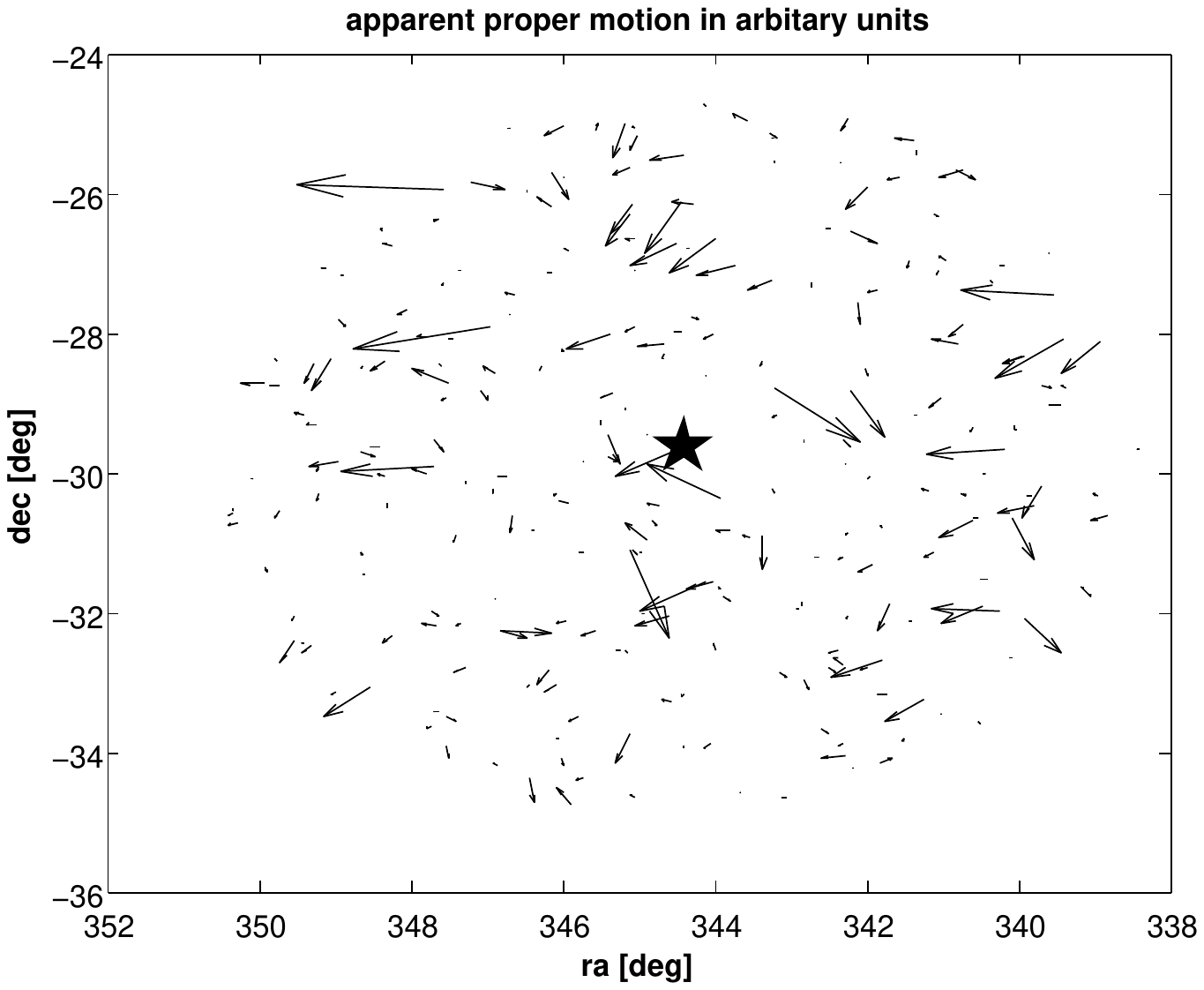}
\caption{{\bf Proper motion} (scaled to arbitrary units) of Fomalhaut's neighbouring stars according 
to the {\it SIMBAD} data base. Fomalhaut is marked by the black solid star in the centre.}
\label{pmF1}
\end{figure}

Formalhaut b is not only located close to a bright star (Formalhaut A), but it also moves with a similar proper motion.
We checked the proper motions of all stars (projected) near Fomalhaut, where this quantity is measured, see Fig. 6.
Many stars move from north-east to south-west (including Fomalhaut), 
suggesting a preferred direction of motion of stars in this field. 
Therefore, an apparently co-moving object located a few parsec behind Fomalhaut (in the same Galactic spiral arm) is well possible.

\subsubsection{Gamma-ray detection~?}

Before we conclude let us also check whether the object might have been detected by some $\gamma$-ray detector.
We have cross-correlated the position of Fomalhaut b with all sources from BATSE and Fermi.

We found two positive possible correlations: \\
Fomalhaut b is located in the positional error ellipse of
a BATSE $\gamma$-ray burst (GRB) source, namely $0.34^{\circ}$ off the BATSE source 11591a at 2000-02-17T06:54:21,
which has a large positional error of $9.4^{\circ}$ and a very low flux of 0.106 photons cm$^{-2}$ s${-1}$
(Stern et al. 2001). \\
Fomalhaut b is also located $0.96^{\circ}$ off the BATSE source named 951022.99 inside its 
large positional error ellipse ($11.3^{\circ}$); this source did not even lead to a follow-up trigger
(Kommers et al. 2001). \\
Given the large error bars of these two BATSE $\gamma$-sources, it is unlikely that Fomalhaut b is related to any of them.

Then, we have retrieved the FERMI LAT data ourselves to search for a source near Fomalhaut.
We used the data up to $15^{\circ}$ around Fomalhaut with a resolution of $0.1^{\circ}$ per pixel in
the energy range from 100 to 300,000 MeV. We detected only the two known sources 2FGL J2258.9-2759 
(separation $97.8^{\prime}$) und 2FGL J2250.8-2808 (separation $126.5^{\prime}$) near Fomalhaut.
Then, we have subtracted those sources from the data, to search again for a faint remaining source,
but we could not detect anything near Fomalhaut.

There is no Fermi source or even Fermi pulsar anywhere near Fomalhaut b.

\section{Conclusions}

The faint object near Formalhaut A (called Fomalhaut~b) remains the subject of intense discussion. 
If one assumes that Fomalhaut~b is gravitationally bound to Fomalhaut~A, 
then the most likely hypothesis seems to be an expanding dust cloud without a central source of $\geq$10\,M$_{Earth}$. 
We show that the body of observations (optical photometry, proper motion, X-ray non-detection) can in principal be fit with a NS. 
In particular, we can explain the SED in the optical wavelength range and the non-detections in the near- and mid-infrared:
The available photometry allows good fits for a blackbody with temperature range from 112,000 to 126,500 K 
(of the optical emitting area)
-- for a neutron star,
this temperature range would yield a distance of $\sim 11$ pc to remain undetected in X-rays as observed. While this may appear to be a fine-tuned
parameter range, one should also keep in mind that such parameters are consistent with {\em all} observables,
while the planetary interpretation has problems with two observational issues (ring-crossing orbit and non-detection in IR).

We also show that it is not unlikely to find one such faint, but unrelated object (or even one of $\sim 10^{8}$ Galactic NSs)
near one of the many bright stars surveyed with deep imaging for planets.
If Fomalhaut b is a neutron star rather than a planet, then the eccentricity of the dust ring around Fomalhaut A might
be explained by either one or more lower-mass, as yet undetected planet(s) or possibly by an eccentric stellar companion.

Given our rough distance estimate for Fomalhaut b as neutron star, 
it might also be possible that Fomalhaut b is both a neutron star and a companion in the Fomalhaut system ($\sim 8$ pc),
i.e. that they orbit around each other - Fomalhaut b as neutron star currently located some tens to hundreds of au behind 
(or before) Fomalhaut A. It should be less problematic for a neutron star component to orbit on an eccentric and/or inclined
orbit (than for a planetary companion).
In such a case, one might expect Bondi-Hoyle accretion of circumstellar material onto the NS and, hence, variable brightness, 
but such a variable brightness would be hard to detect for an object as faint as $\sim 25$ mag with large error bars 
(see Table 1 and Figs. 4 \& 5).

If the companion candidate to Fomalhaut is indeed a NS at some 11 pc distance, it would be the clostest known NS.
With a total of some $10^{8}$ NSs in the Galaxy ($\sim 30$ kpc diameter, $\sim 0.6$ kpc thickness),
we would expect $\sim 2.2$ NSs between 8 and 14 pc (i.e. $\pm 3$ pc around the best fit, the lower value being the lowest
allowed distance). We may have noticed one of them.
If we would restrict this estimate to young (self-luminous) NSs, the expected number would be smaller.
If the companion candidate to Fomalhaut is indeed a NS, then it may either be an exception
(as self-luminous very nearby NS) -- or it may be an old NS re-heated by accretion from the interstellar material.
For the latter, this NS would need to travel with small velocity $v$ through the interstellar material 
(Bondi-Hoyle accretion scales with $v^{-3}$), which is indeed the case: $\sim 30$~km/s only as two-dimensional
velocity from its proper motion at $\sim 11$ pc, respectively, which is slow for NSs.

Our NS hypothesis could be tested by observations in the UV (e.g. with the ACS Solar Blind Camera):
If the object is detected and shows a similar flat SED as in the optical, 
than a planet or dust cloud could be ruled out and a background neutron star becomes the most likely explanation.
If radio, X-ray, or gamma-ray pulsations (in the typical range as for NSs, i.e. few milli-seconds to $\sim 20$ seconds) 
could be detected in the Fomalhaut companion candidate in very deep observations,
it would certainly be a neutron star (we would expect pulsations around $\sim 1$ to $\sim 10$ seconds, 
rather than below one second, because it would need to be a middle-aged to old neutron star).
An X-ray detection would also yield a more stringent constraint on the age.
A detection of gravitational waves due to rotation of a non-spherical object would confirm its compactness.

\section*{Acknowledgments}
RN, MMH and JGS acknowledges support by the Deutsche Forschungsgemeinschaft (DFG) through SFB/TR 7 ``Gravitationswellenastronomie''. 
We would like to thank Alexander Krivov and Markus Mugrauer for valueable comments on an earlier version of this manuscript.
CG wishes to acknowledge Deutsche Forschungsgemeinschaft (DFG) for grant MU 2695/ 13-1. 
We obtained the AMES models from phoenix.ens-lyon.fr/Grids/AMES-Cond/SPECTRA.
This research has made use of the SIMBAD database, operated at CDS, Strasbourg, France. 
This research has made use of NASA's Astrophysics Data System Bibliographic Services. 
We express special thanks to Donna Keeley for language editing of the paper.
We also would like to thank an anonymous referee for very good comments.

\end{document}